\documentclass{emulateapj}
\newcommand{\eV}{\mbox{ eV}}
\newcommand{\kel}{\mbox{ K}}

\newcommand{\secinv}{\mbox{ s$^{-1}$}}

\newcommand{\hunits}{\mbox{ km s$^{-1}$ Mpc$^{-1}$}}

\newcommand{\recunits}{\mbox{ cm$^{3}$ s$^{-1}$}}

\newcommand{\lya}{Ly$\alpha$ }
\newcommand{\lyans}{Ly$\alpha$} 

\newcommand{\deriv}{{\rm d}}
\newcommand{\bq}{\begin{equation}}
\newcommand{\eq}{\end{equation}}
\newcommand{\bqa}{\begin{eqnarray}}
\newcommand{\eqa}{\end{eqnarray}}
\def\VEV#1{\left\langle #1\right\rangle} 

\begin{document}

\title{The Equation of State of the Intergalactic Medium After Hydrogen Reionization}

\author{Steven R.  Furlanetto\altaffilmark{1} \& S.~Peng Oh\altaffilmark{2}}

\altaffiltext{1} {Department of Physics and Astronomy, University of California, Los Angeles, CA 90095, USA; sfurlane@astro.ucla.edu}

\altaffiltext{2} {Department of Physics, University of California, Santa Barbara, CA 93106, USA}

\begin{abstract}
We use an analytic model to study how inhomogeneous hydrogen reionization affects the temperature distribution of the intergalactic medium (IGM).  During this process, the residual energy of each ionizing photon is deposited in the IGM as heat, increasing its temperature to $20,000$--$30,000 \kel$; subsequent expansion of the Universe then cools the gas.  Because reionization most likely proceeds from high to low densities, underdense voids are ionized last, have less time to cool, and are (on average) warmer than mean-density gas immediately after reionization is complete (an ``inverted" equation of state).  From this initial configuration, the low-density gas cools quickly and eventually returns to a more normal equation of state.  The rapidly evolving temperature introduces systematic uncertainties in measurements of the ionizing background at $z \sim 6$.  For example, late reionization implies rapid cooling, so that the ionizing background would have to evolve even more rapidly at $z \sim 5$--6 than typically claimed.  This degeneracy is difficult to disentangle, because the \lya forest probes only a narrow range in densities (over which the gas is nearly isothermal).  However, higher Lyman-series transitions probe wider density ranges, sampling different effective temperatures, and offer a new way to measure the IGM equation of state that should work where nearly saturated absorption precludes other methods. This will help to separate evolution in temperature from that in the ionizing background.  While more detailed study with hydrodynamic simulations is needed, we show that such measurements could potentially distinguish early and late reionization using only a handful of lines of sight.
\end{abstract}
  

\section{Introduction} \label{intro}

The two most dramatic events in the history of the intergalactic medium (IGM) are the reionization of hydrogen (by the first generations of galaxies) and helium (by quasars).  These have become key landmarks for both observational and theoretical cosmologists in the past several years.  Evidence for hydrogen reionization comes from a number of directions, none of them clear but all consistent with (possibly extended) reionization at $z \sim 6$--$10$ (see \citealt{fan06-review, furl06-review} for recent reviews).  Helium reionization is thought to occur during the quasar era, at $z \sim 3$, with a wide variety of supporting evidence -- though much of it is controversial, and a clear picture has yet to emerge (see \citealt{furl08-helium} for a recent summary of the observations).

These two reionization epochs are largely responsible for determining the thermal history of the IGM.  Before hydrogen reionization, the neutral IGM cooled adiabatically until the first structures formed, probably reaching temperatures $T \la 10 \kel$.  X-rays from the first galaxies most likely slowly heated the neutral IGM to $T \la 1000 \kel$ \citep{oh01, venkatesan01, kuhlen06-21cm, furl06-glob}.  However, hydrogen reionization caused a much more dramatic change:  the $\sim 5$--$10 \eV$ leftover from each ionizing photon heated the IGM to $\sim 2$--$3 \times 10^4 \kel$ \citep{miralda94, abel99, tittley07, trac08}.  The harder photons responsible for helium reionization could have reheated the IGM to similar, or even larger, temperatures \citep{hui97, furl08-igmtemp, mcquinn08, bolton08}.  

Once reionization is complete, this heating channel slows dramatically -- because only the relatively small fraction of ions that recombine couple to the photoionizing background.  The subsequent temperature evolution is determined primarily by a balance of adiabatic heating and/or cooling, photo-heating, and Compton cooling \citep{miralda94, hui97}.  The competition between these processes forces the gas temperature to approach an asymptotic form set by the background ionizing spectrum \citep{hui97, hui03}.  Because the magnitude (and indeed sign) of the adiabatic term depends on whether the gas is over- or underdense, the IGM assumes an equation of state $T \approx T_0 (1 + \delta)^{\gamma-1}$, where $\delta$ is the fractional overdensity of the gas element and $\gamma > 0$ is nearly independent of $\delta$ in most simple models.  

This equation of state inevitably affects many observables of the \lya forest.  In particular, the observed temperature evolution has been used to constrain the epochs of helium and hydrogen reionization:  most dramatically, the velocity widths of \lya forest absorbers seem to increase sharply at $z \sim 3.2$ \citep{schaye00}, while the equation of state simultaneously flattens \citep{schaye00, ricotti00}.  This may be a result of helium reionization (though see \citealt{mcdonald01}).  Recent models of inhomogeneous helium reionization show that these features are consistent with the behavior expected near that event \citep{gleser05, furl08-igmtemp, mcquinn08}.  Moreover, these models showed that helium reionization could dramatically transform the IGM thermal structure:  \emph{if} low-density voids are ionized last, they suffer the least amount of cooling and hence contain the hottest gas at the end of reionization, leading to an ``inverted" equation of state.  Observations now suggest that something similar may indeed be occurring at $z \sim 3$ \citep{becker07, bolton08-eos}.  However, helium reionization is probably driven by rare, bright quasars; in that case, the effects are less dramatic because the nearly-random distribution of sources makes the process more stochastic, and the equation of state may flatten but not invert \citep{furl08-helium, mcquinn08, bolton08}.

Hydrogen reionization is an even more dramatic episode in the thermal history of the IGM, because it increases the IGM temperature by at least an order of magnitude.  Moreover, it is most likely driven by large numbers of small galaxies, so the ionization topology (and hence thermal structure) are driven largely by the underlying density field \citep{barkana04, furl04-bub}.  Unfortunately, helium reionization erases any information from this phase, so it is much more difficult to probe.  One approach is to use the direct temperature measurements at $z \ga 4$ as a ``fossil" record of hydrogen reionization.  The inferred IGM temperature is relatively large, so this appears to require reionization at $z \la 10$ \citep{theuns02-reion, hui03}.   However, because the IGM quickly approaches the aforementioned asymptotic equation of state, such inferences are difficult.

Instead, it would be better to study the thermal effects of reionization soon after that era ends, when the cooling is most rapid \citep{trac08}.  This is, unfortunately, a difficult proposition, because the \lya forest -- our principal tool to study the IGM -- becomes nearly saturated in absorption at $z \ga 5$ (e.g., \citealt{fan01, fan06}).  Thus no serious attempt has been made to measure the thermal structure at these redshifts.  Nevertheless, it is crucial to interpreting the observations:  the (temperature-dependent) recombination rate $\alpha(T)$ and the amplitude of the ionizing background $\Gamma$ affect the (observed) optical depth only in the combination $\Gamma/\alpha(T)$, so independent estimates (or robust priors) on the temperature are required to accurately constrain the growing ionizing background at $z \sim 6$ \citep{fan02, fan06, bolton07}.  The rapid cooling expected after reionization ends may have serious implications for our understanding of this epoch.  Moreover, recent simulations show that the equation of state may be exceedingly complex shortly after reionization -- with both an inversion and a large amount of scatter \citep{bolton04, trac08}.  By contrast, almost all studies constraining $\Gamma$ assume an isothermal IGM with no temperature evolution, or marginalize over simple equations of state.  

Indeed, if we could measure the evolution of the equation of state as a function of redshift, we could place constraints on both the topology and redshift of reionization. The initial conditions (particularly the presence of an inverted equation of state) imprint information about the reionization topology, while the rate of evolution -- which is most rapid immediately after reionization -- tells us when that event occurred.  Unfortunately, the saturation of the Ly$\alpha$ forest, as well as the paucity of data at $z \sim 6$, imply that conventional techniques to measure the equation of state fail at high redshift. In this paper, we suggest that the flux transmission ratios of \ion{H}{1} Ly$\alpha$ and Ly$\beta$, which probe different density regimes, can be used to measure the equation of state, provided the ionizing background does not vary systematically as a function of overdensity. This may open up studies of the equation of state at hitherto inaccessible redshifts, allowing us to probe its evolution over cosmic time. 

In this paper, we apply our model of inhomogeneous reionization (fully detailed in \citealt{furl08-igmtemp}, hereafter FO08) to this important event.  We will study how reionization affects the thermal structure of the IGM, and through that the \lya forest.  We briefly describe the model in \S \ref{model-therm}.  We examine the resulting thermal histories in \S \ref{temphist} and their effects on the \lya forest in \S \ref{lyaforest}. We discuss the possibility of measuring the evolution of the equation of state in \S \ref{measure}. We conclude in \S \ref{disc}.

In our numerical calculations, we assume a cosmology with $\Omega_m=0.26$, $\Omega_\Lambda=0.74$, $\Omega_b=0.044$, $H=100 h \hunits$ (with $h=0.74$), $n=0.95$, and $\sigma_8=0.8$, consistent with the most recent measurements \citep{dunkley08}.  Unless otherwise specified, we use comoving units for all distances.

\section{Model for the Thermal History} \label{model-therm}

Our model for inhomogeneous reionization is described fully in FO08.\footnote{The original paper focused on helium reionization; here we only consider hydrogen.}  In brief, it is based on the excursion set approach to reionization of \citet{furl04-bub}, which allows us to calculate the distribution of reionization redshifts for gas parcels as a function of their density.  Given a final reionization redshift $z_r$, the history is determined by the assumed ionizing efficiencies of dark matter halos.  Our fiducial model takes a constant ionizing efficiency in all halos with $T_{\rm vir} > 10^4 \kel$ (the threshold for atomic cooling in a hydrogen-helium gas), typical of many reionization models in the literature (see, e.g., \citealt{barkana01}).  

The excursion set model essentially evaluates the probability that a small gas parcel of a specified density will lie inside an ionized bubble during various stages of reionization (and hence determines the local reionization redshift for a given IGM patch).  It implicitly assumes that the ionized bubbles trace dense regions surrounding the ionizing sources, so that dense gas elements are ionized relatively early, while voids must await the final stages of reionization.  Note that, in FO08, we allowed for the possibility of ``stochastic" reionization where the ionization history of a given gas parcel was independent of its density.  While this is important for helium reionization, because that event is driven by rare and bright quasars \citep{furl08-helium, mcquinn08}, such a stochastic phase is probably not relevant during hydrogen reionization, which is driven by a multitude of faint galaxies.  We will therefore focus exclusively on the density-driven model.

\begin{figure*}
\plottwo{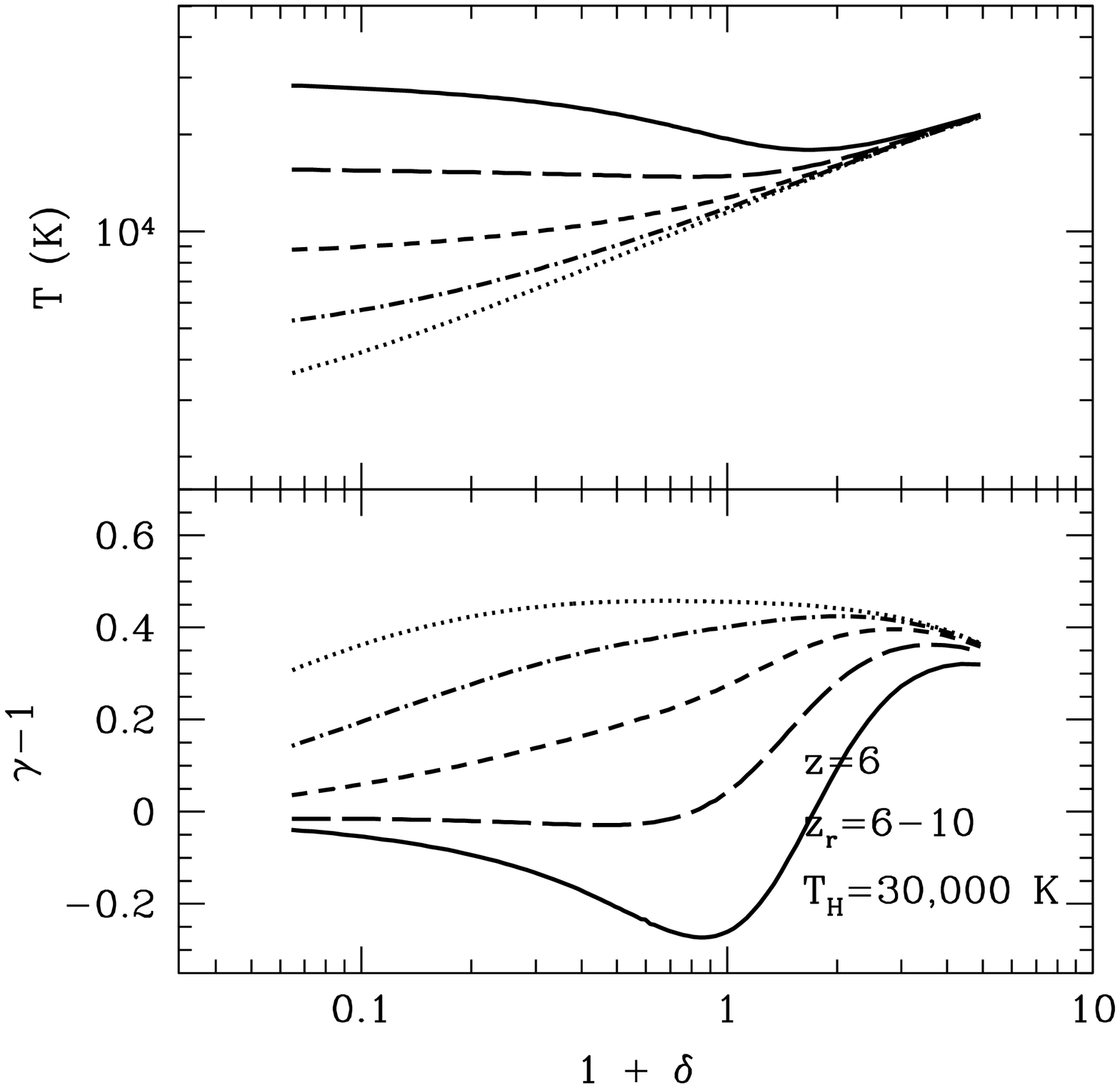}{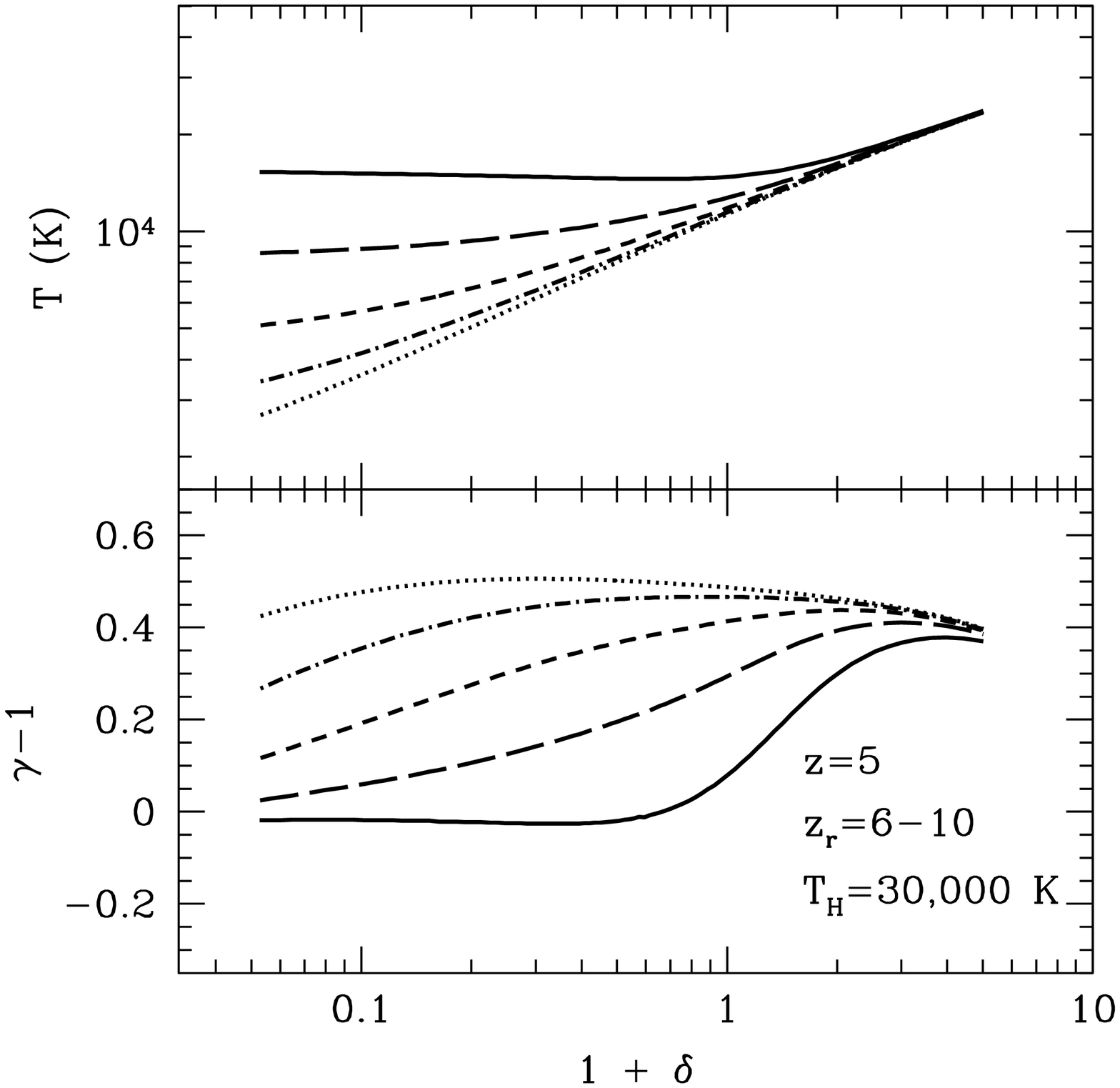}
\caption{\emph{Top:}  Median temperature at $z=6$ and $z=5$ (left and right panels, respectively) as a function of density after hydrogen reionization.  The solid, long-dashed, short-dashed, dot-dashed, and dotted curves assume that reionization completes at $z_r=6,\,7,\,8,\,9,$ and $10$. \emph{Bottom:}  Local equation of state index, $\gamma-1$, for the same scenarios.  All panels assume that the initial post-reionization temperature is $T_H=30,000 \kel$.  }
\label{fig:TH30}
\end{figure*}

Once the reionization history of an IGM parcel is chosen, we follow the subsequent thermal evolution using a simplified version of \citet{hui97}, which includes all the relevant atomic heating, recombination, and photoionization processes as well as the density evolution (traced through the spherical collapse model; see FO08 for details).\footnote{For computational simplicity, once a parcel is ionized we assume that it remains in ionization equilibrium with the radiation background.  We have tested this assumption against a non-equilibrium code (see FO08) and find it to be an excellent approximation in all cases.}  After reionization, the dominant processes are photoionization heating and adiabatic cooling (with Compton cooling an important factor as well at the highest redshifts).

\subsection{The Initial Post-Reionization Temperature} \label{Tinit}

By construction, we cannot use our code while a gas parcel is first being ionized.  It is during this brief period that the gas is rapidly heated by the leftover energy of each ionizing photon.  We instead set the initial post-reionization temperature, $T_H$, by hand.  A number of factors that cannot self-consistently be included in our code affect this initial temperature -- most importantly, we do not know the radiation spectrum that illuminates each gas element.  This obviously depends on the UV spectra of the star-forming galaxies that provide the ionizing photons.  The Starburst99 models show that low-metallicity galaxies have a luminosity density that varies roughly as $L_\nu \propto \nu^{-2}$ between the hydrogen and helium ionization edges (Fig. 2e, \citealt{leitherer99}).  However, other factors are also involved -- for example, spectral filtering by dense gas parcels (which may absorb low-energy photons but allow high-energy photons to pass) makes the background spectrum inhomogeneous and harder than the input spectrum \citep{abel99}. 

We will therefore appeal to some general arguments that bound the likely value of $T_H$ \citep{miralda94, abel99}.  If the region between the gas element and the ionizing sources is optically thin, the mean excess energy of ionizing photons $E_{\rm thin}$ is the average of the entire spectrum weighted by the ionization cross section, $\sigma_i \propto E^{-3}$, so low-metallicity stars yield $\VEV{E_{\rm thin}}/E_i \approx 1/4$, where $E_i$ is the ionization potential.\footnote{Here we have ignored absorption by neutral helium and included all photons with $13.6 \eV < E < 54.4 \eV$; in reality, $\sim 1/8$ of photons above 24.6 eV will ionize helium instead, producing less heat than our standard expression.  But this provides only a small correction to our already-uncertain estimates.}  However, the hydrogen density is so large that even most low-density gas systems will themselves be optically thick.  In that limit, where all ionizing photons are absorbed, we do not weight by the cross section, so $\VEV{E_{\rm thick}}/E_i \approx 3/5$.  In either case, this energy must then be shared with all the IGM baryons  through Coulomb interactions; the net temperature change is then $\Delta T \approx 0.5 (2/3k_B)  \VEV{E} \sim 30,000 \kel$ for the optically thick case.  In the following, we assume that $T_H = \Delta T$, which should be an excellent approximation given the low IGM temperatures expected before reionization, even in the presence of X-ray heating (e.g., \citealt{furl06-glob}).\footnote{Note that FO08 assumed a significantly lower post-hydrogen reionization temperature, $T_H=15,000 \kel$, and so found much weaker effects at $z \sim 4$ than we do here.}

\begin{figure*}
\plottwo{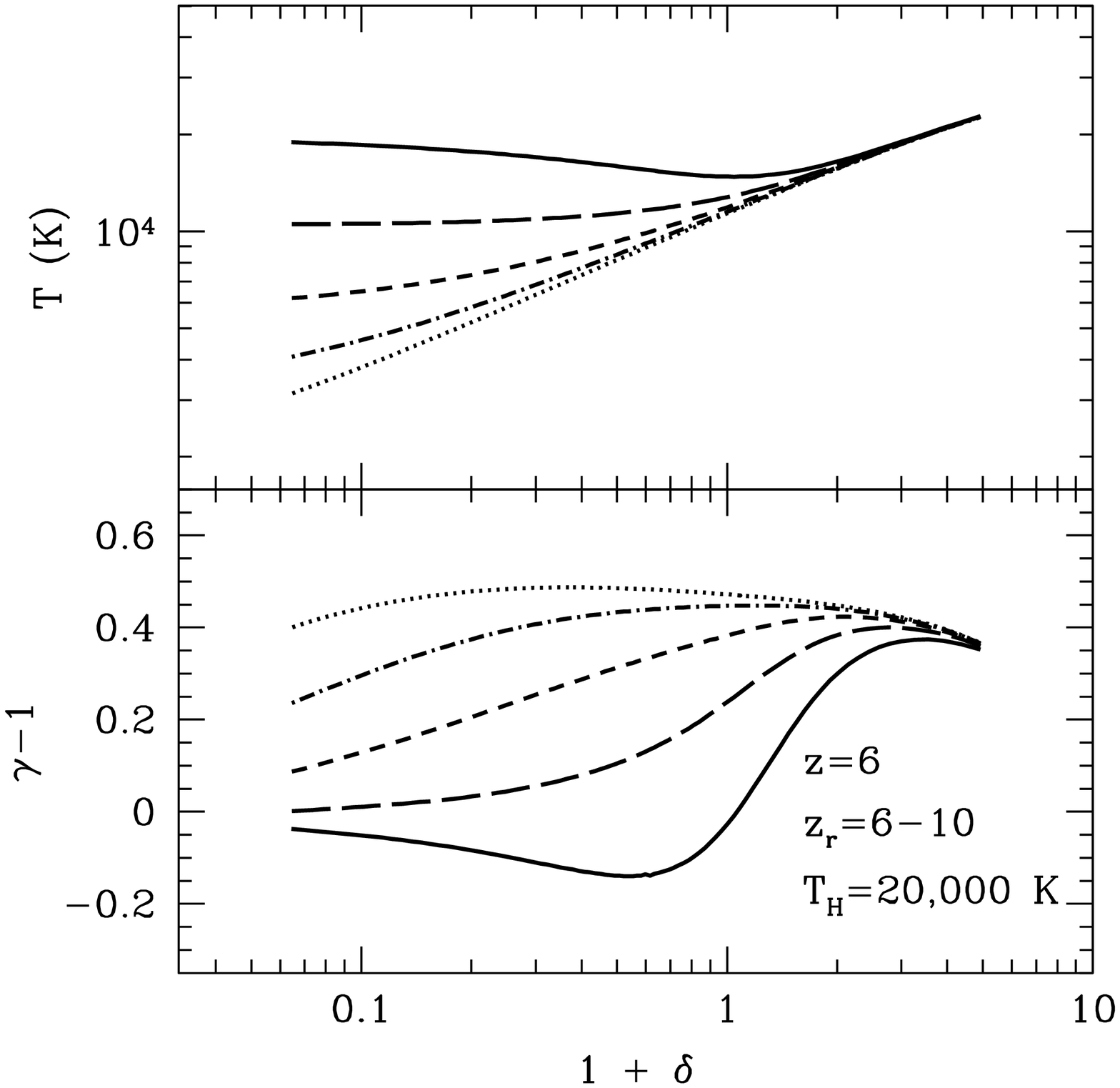}{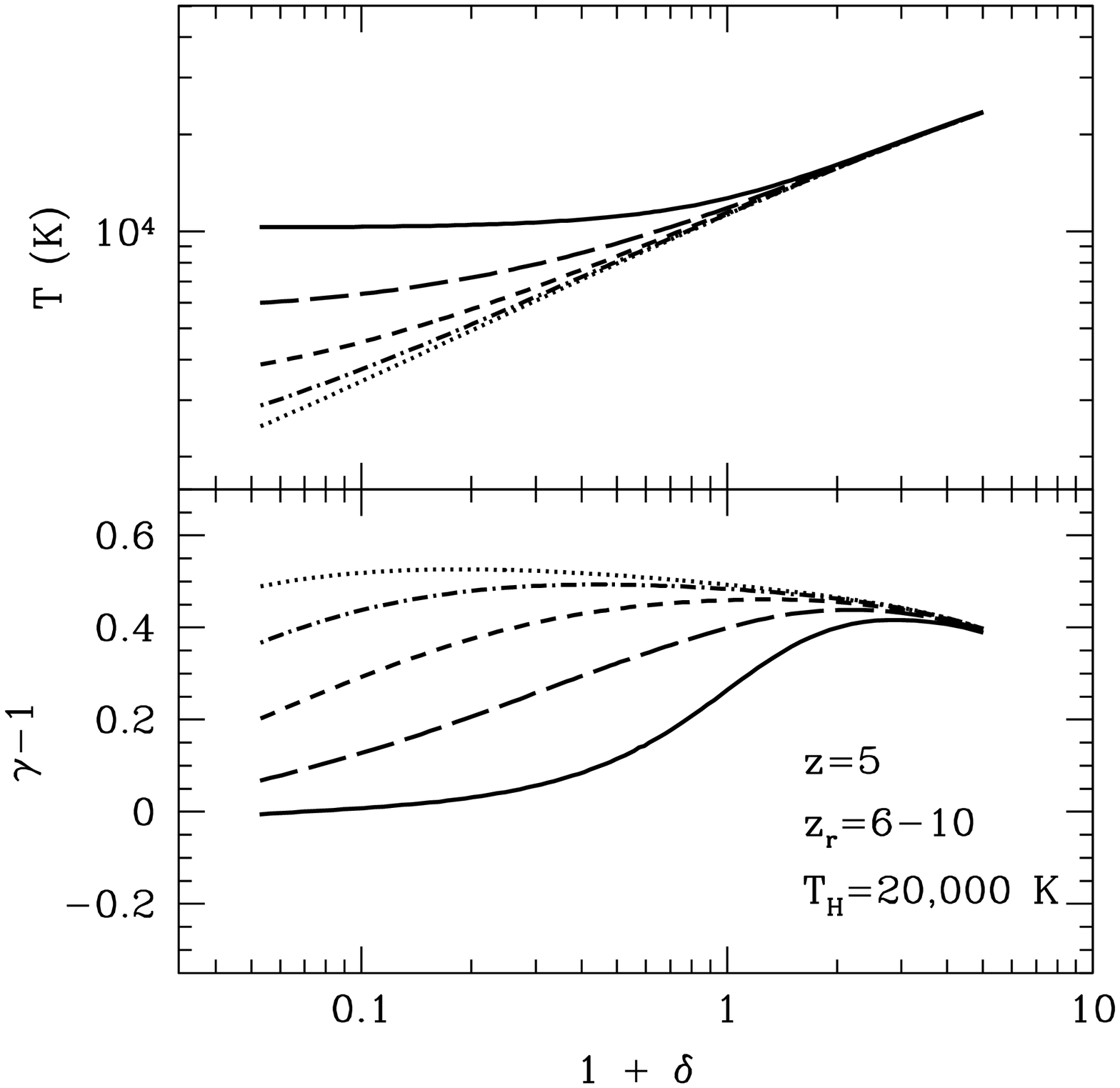}
\caption{As Fig.~\ref{fig:TH30}, but for $T_H=20,000 \kel$.  }
\label{fig:TH20}
\end{figure*}

Our code also requires that the ionizing background be specified once a gas parcel is ionized.  For simplicity, we will take a constant power-law shape and amplitude at all redshifts, with the angle-averaged specific intensity $J_\nu \propto \nu^{-\alpha}$ from 13.6 to 54.4 eV, and $\alpha=1.5$ (corresponding to a slightly hardened stellar spectrum).  In our calculations, we set the ionization rate $\Gamma = 10^{-12} \Gamma_{12} \secinv$ to a relatively high value that is appropriate well after reionization, $\Gamma_{12}=1$.  Fortunately, these assumptions make almost no difference to our results.  Although we do not allow the amplitude of the ionizing background to evolve (of course it actually increases as more and more of the IGM is ionized), this actually has no direct effect on the thermal history:  a larger $\Gamma$ increases the rate at which neutral atoms interact with the background, but it reduces the neutral fraction by exactly the same amount.  It only weakly affects secondary heating and cooling mechanisms, such as Compton scattering, by slightly increasing the density of free electrons.  In addition, note that, after the initial reionization, the shape of the ionizing background also has little effect on the temperatures. 

\section{The Temperature History} \label{temphist}

The top panels of Figure~\ref{fig:TH30} show the median IGM temperature as a function of density (here $\rho/\bar{\rho}=1+\delta$) for gas at $z=6$ and 5 (left and right panels, respectively), if reionization occurs at $z_r=6,\,7,\,8,\,9,$ and 10 (from top to bottom in each panel) if $T_H=30,000 \kel$.  The bottom panels show the effective equation of state at each density, 
\bq
\gamma - 1 \equiv {\deriv \log T \over \deriv \log \delta}.
\label{eq:gammadefn}
\eq
Figure~\ref{fig:TH20} shows similar information for models with $T_H=20,000 \kel$.

If reionization occurs early enough, the competition between photoheating and adiabatic cooling (including both the expansion of the universe and structure formation, whose effect depends on $\delta$) drives the IGM toward an asymptotic equation of state \citep{hui97, hui03}; at the high redshifts of interest here, this process is accelerated significantly by Compton cooling.  This asymptote is typically described as a simple power law, $T \propto (1+\delta)^{\gamma-1}$, which we find to be a good approximation:  reaching $(\gamma-1) \approx 0.6$ long after reionization, with some flattening at very high densities.  However, at redshifts closer to the time of reionization, a simple power law obviously no longer suffices to describe the results.  Although $\gamma$ remains nearly the same at high densities, there is substantial flattening at lower densities -- and, sufficiently close to reionization, it even becomes inverted ($\gamma-1<0$) near and below the mean density.  

This occurs because the highest-density regions -- which preferentially lie near galaxies -- are ionized first, while the lowest-density voids are far from sources and so get ionized last.  All are initially heated to the same temperature ($T_H$), but the former have longer to cool.  (The highest density regions undergo rapid enough adiabatic heating that they stay warm.)  Thus, when reionization ends, the hottest regions correspond to the most-recently ionized voids.  Recent numerical simulations of reionization by \citet{trac08} show a very similar median relation (although more similar to our Fig.~\ref{fig:TH20} than the higher $T_H$ case).  An inverted equation of state could also occur during helium reionization, if small sources are common enough (FO08), although in practice it appears unlikely given the dominance of bright quasars and the more complex radiative transfer \citep{mcquinn08, bolton08}.

\begin{figure*}
\plottwo{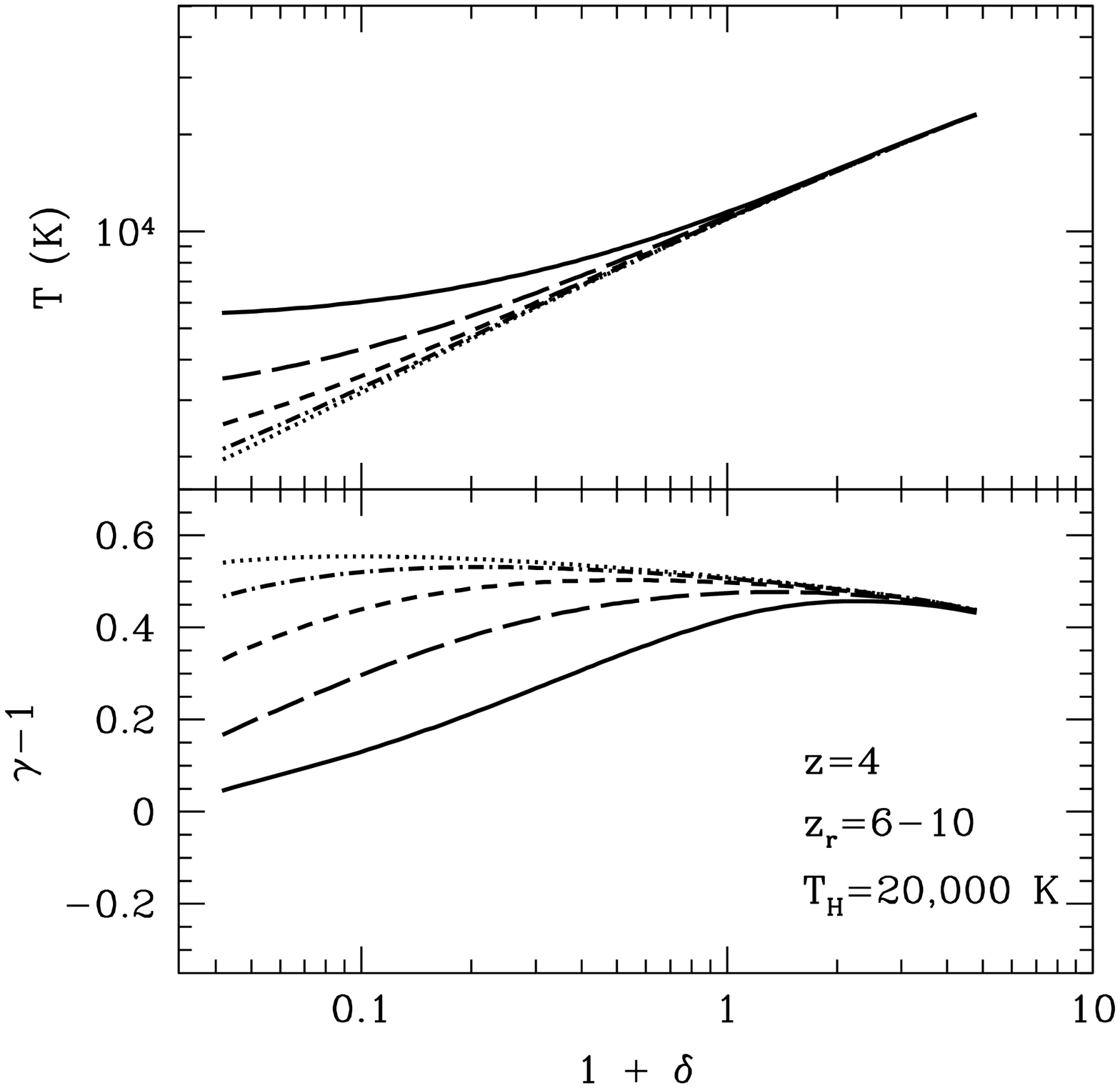}{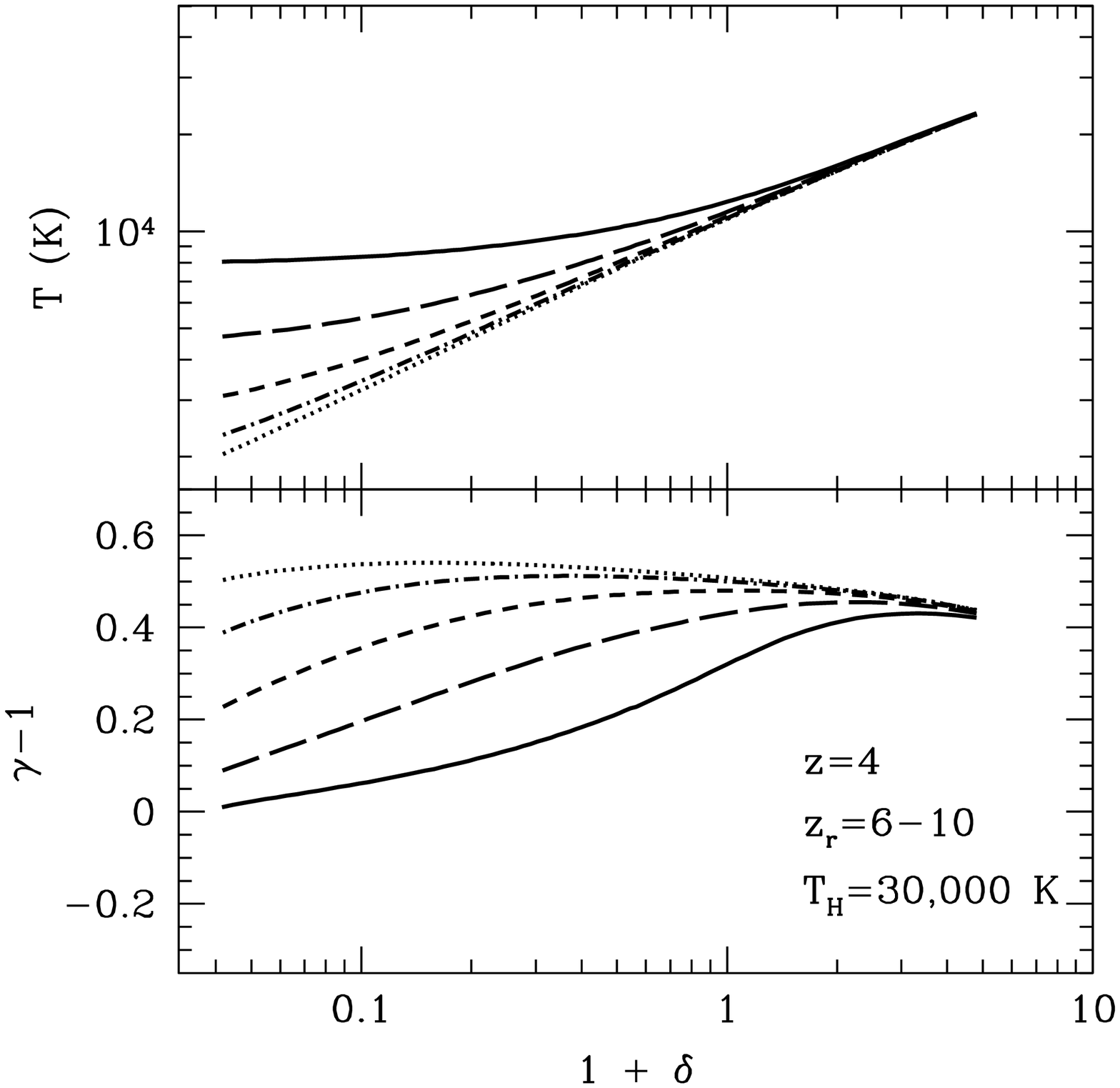}
\caption{As Fig.~\ref{fig:TH20}, but at $z=4$ and for $T_H=20,000 \kel$ (left panels) and $30,000 \kel$ (right panels).  }
\label{fig:z4}
\end{figure*}

The IGM cools rapidly after this (especially with the addition of Compton cooling at high redshifts), and most quickly in the voids, so this inversion does not last long.  Once $z \la z_r - 1$, the low density gas has become nearly isothermal, from there slowly returning to a monotonic equation of state and eventually the thermal asymptote.  Thus, photoheating will only substantially affect the equation of state for a relatively brief period after hydrogen reionization.  In particular, Figure~\ref{fig:z4} shows that the effects will be relatively modest by $z=4$:  the differences at the mean density are only $\sim 2000 \kel$, although they remain a factor of a few at the lowest densities.

There have been attempts to constrain $z_r$ from the \lya forest temperature at $z \sim 4$ \citep{theuns02-reion, hui03}, under the assumption that the temperature is uncontaminated by helium reionization there.  Unfortunately, the data are not nearly good enough to measure the relatively small expected difference.  In fact, temperatures inferred from the existing data generally lie well above predictions that include only hydrogen photoheating \citep{schaye00, zald01}.

\begin{figure*}
\plottwo{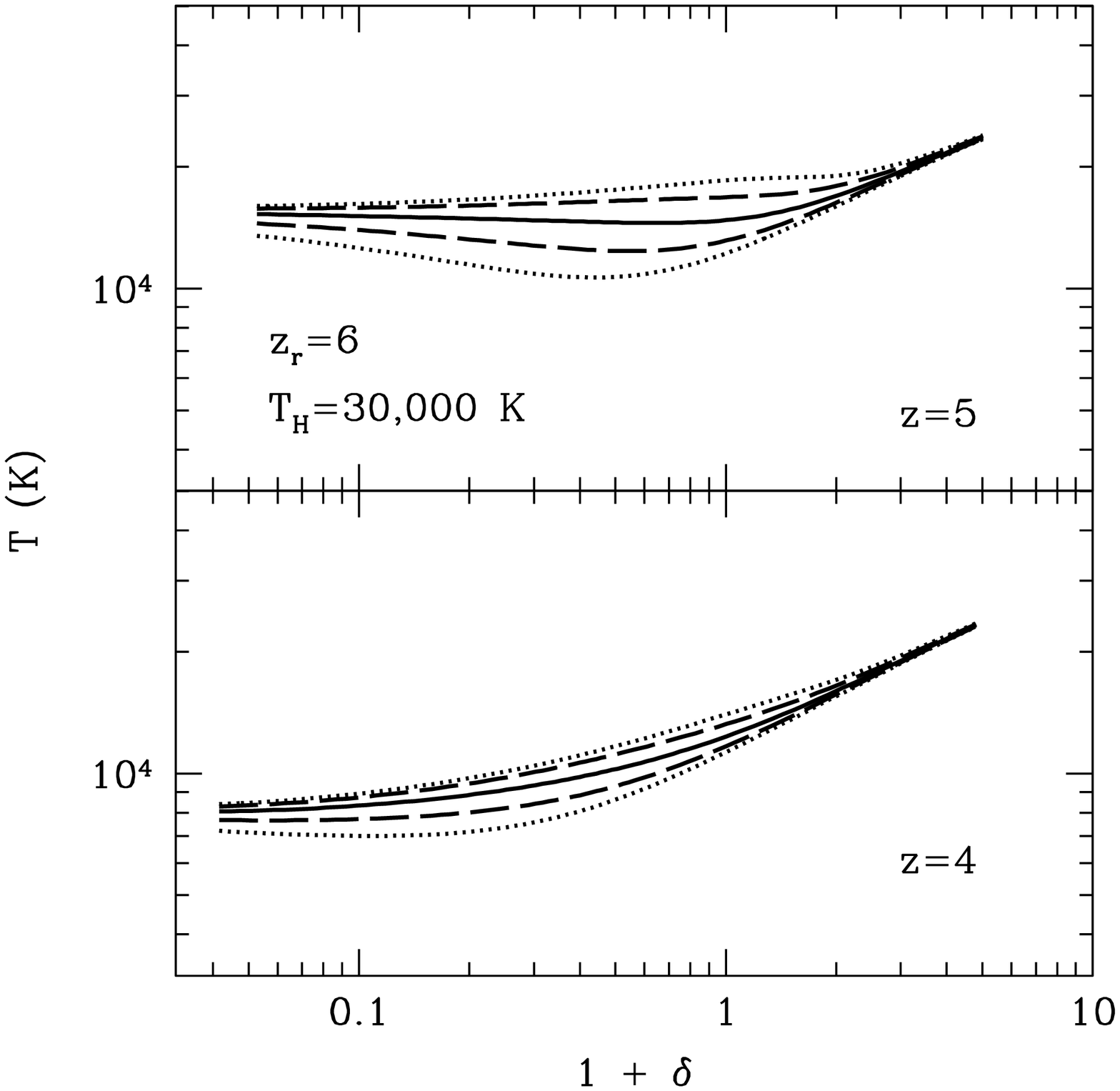}{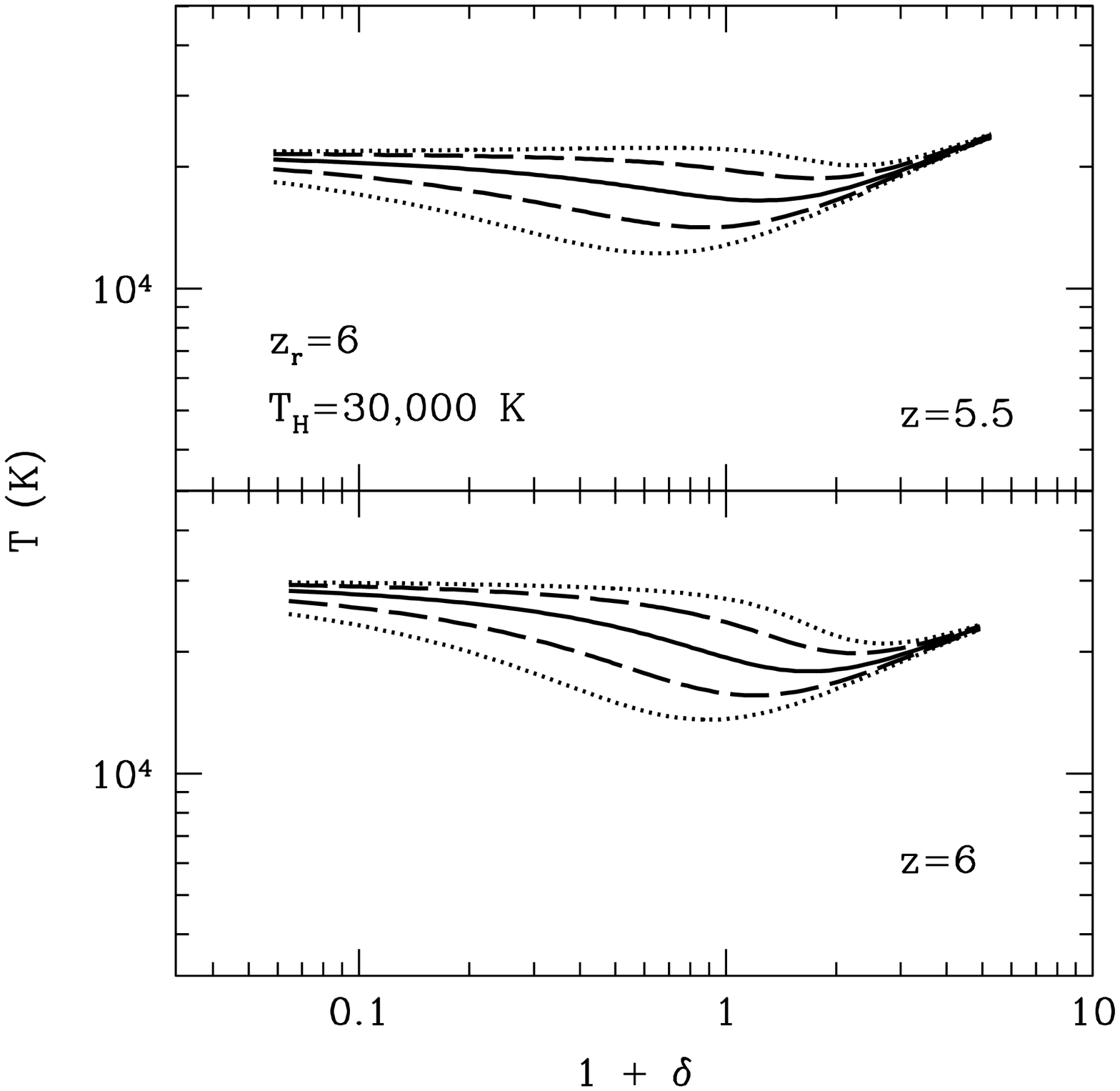}
\caption{Distribution of IGM temperature as a function of density, during and after reionization.  All panels assume $z_r=6$ and take $z=4,\,5,\,5.5$, and 6, clockwise from bottom left.  The curves show the 10th, 25th, 50th, 75th, and 90th percentiles of the distributions.  Both panels assume that the post-reionization temperature is $T_H=30,000 \kel$.  }
\label{fig:spread}
\end{figure*}

Figure~\ref{fig:spread} shows the second major effect of inhomogeneous reionization on the IGM temperature distribution:  because different gas parcels (even at the same local density) have different ionization histories, reionization creates a significant scatter in the temperature distribution.  The curves in each panel show the 10th, 25th, 50th, 75th, and 90th percentiles of the distributions, from bottom to top; the different panels show different times after reionization, all assuming $z_r=6$.  

Again, because reionization proceeds ``inside-out," the highest density regions are ionized first, with relatively little scatter in their reionization redshifts (which is then decreased even further by adiabatic heating).  The lowest density regions are ionized last, again with relatively little scatter:  in the excursion set formalism, such voids are always near the last to be ionized -- in this case at $6 \la z \la 6.5$ -- so the final spread in temperatures is modest.  However, the probability that gas near the mean density has been ionized at any time in the central period of reionization is roughly equal to the mean ionized fraction, so there is a considerable spread in the temperatures around $\delta \sim 0$.  The factor of two spread persists until $z \sim 5$, but the relation then tightens as gas approaches the thermal asymptote, and by $z=4$ the scatter is small.  In principle, this reduction in scatter could be a tell-tale signature of when reionization took place, as has been suggested for helium reionization \citep{hui03}, although that has not yet been detected (e.g., \citealt{theuns02-temp, zald02}). For hydrogen reionization, it is unfortunately difficult to obtain sufficient statistics to measure the scatter in the equation of state at high redshifts, while the lower redshift observations at $z\sim 4$ are likely contaminated by helium reionization. 

The reionization simulations of \citet{trac08} show comparable scatter near the mean density but much more in low-density voids and along the high-density tail.   This suggests that our excursion set model relating the large-scale ionized regions to small gas parcels may contain too much of a correlation across scales, although there are other complexities in the simulation (in particular, differing source prescriptions, radiative transfer, and the complex geometry of the cosmic web) that can explain these differences.  In any case, increased scatter is likely to strengthen the observable consequences that we describe below.

\section{Effects on the HI \lya Forest} \label{lyaforest}

We have seen in the previous section that the equation of state evolves significantly after reionization completes.  Here we will consider the implications of this phenomenon for observables, and in particular the HI \lya forest.  During the helium reionization era, the corresponding effects are substantial:  the temperature increase and flattening of the equation of state may have already been observed directly, through the line width distribution in the \lya forest (\citealt{schaye99, schaye00, ricotti00}, though see also \citealt{mcdonald01}), and indirectly, possibly through the evolving effective optical depth \citep{bernardi03, faucher08}.  The observed behavior matches the qualitative expectations of theoretical studies of helium reionization, although not necessarily the details -- especially for the optical depth \citep{theuns02-sdss, gleser05, furl08-igmtemp, mcquinn08, bolton09}.

However, extracting the details of the equation of state following hydrogen reionization promises to be much more difficult.  Direct methods, which rely on the linewidths of the features, are useless because the forest is so saturated -- with individual features no longer identifiable.  Instead, we will likely need to constrain the thermal evolution indirectly, through the overall transmission in the forest.

To examine how the optical depth evolves, we follow the model for the IGM density structure of \citet{miralda00}.  They found that a reasonable fit to the volume-averaged density distribution of IGM gas, $P_V(\Delta)$ (where $\Delta=1+\delta$), in simulations at $z \sim 2$--$4$ is:
\bq
P_V(\Delta) \, \deriv \Delta = A_0 \Delta^{-\beta} \exp \left[ - \frac{(\Delta^{-2/3} - C_0)^2}{2(2 \delta_0/3)^2} \right] \, \deriv \Delta.
\label{eq:pvd}
\eq
Intuitively, the underlying Gaussian density fluctuations are modified through nonlinear void growth and a power law tail at large $\Delta$.  Here $\delta_0$ essentially represents the variance of density fluctuations smoothed on the Jeans scale for an ionized medium; thus $\delta_0 \propto (1+z)^{-1}$ at  high redshifts (such as those here).  The power-law exponent $\beta$ determines the behavior at large densities; for isothermal spheres, it is $\beta=2.5$, which we assume to be appropriate at $z=6$.  One caveat is necessary here:  the \citet{miralda00} simulations included Jeans smoothing from a specific reionization history, so we cannot use it to describe the dynamical evolution of the gas distribution that follows reionization \citep{pawlik08}.

We will only consider the most easily measured characteristic of the absorption:  the effective optical depth, $\tau_{\rm eff}$, which simply parameterizes the  mean transmission in \lyans, ${\mathcal T}_\alpha \equiv \exp( - \tau_{{\rm eff},\alpha})$.  We estimate the transmission by simply integrating over $T$ and $\Delta$,
\bq
e^{-\tau_{{\rm eff},\alpha}} \equiv \int \deriv \Delta \, P_V(\Delta) \int \deriv T \, p(T|\Delta) e^{-\tau_\alpha(\Delta, T)},
\label{eq:taueff}
\eq
where $p(T|\Delta)$ is the probability that a gas parcel of density $\Delta$ has a temperature $T$ and is given by our inhomogeneous reionization model.  With the ``fluctuating Gunn-Peterson trough approximation" (e.g., \citealt{gunn65, rauch97, croft98-ps}), the \lya optical depth of a gas parcel is approximately
\bq
\tau_\alpha(\Delta, T) \approx 13 {\Delta^2 T_4^{-0.7} \over \Gamma_{12}} \left( {1+z \over 7} \right)^{9/2},
\label{eq:fgpa}
\eq
where we have used the case-A recombination coefficient, $\alpha_A \approx 4.2 \times 10^{-13} T_4^{-0.7} \recunits$ (appropriate for the low-density diffuse IGM, where most of the transmission will occur), $T_4 = T/(10^4 \kel)$, and we have assumed a large enough redshift that the cosmological constant can be ignored.  Unfortunately, this procedure does not exactly reproduce the features of the \lya forest, and it is necessary to renormalize $\tau_\alpha$ in order to match measurements from more detailed studies of the forest.  Following \citet{fan02}
we introduce a correction factor $\kappa = 0.3$ on the right hand side of equation~(\ref{eq:fgpa}) to reproduce the $\tau_{{\rm eff},\alpha}$--$\Gamma$ relation at $z=4$ found by \citet{mcdonald01-Gamma}.  Although this is a large factor, the hope is that it does not evolve significantly with redshift, so that our results at other redshifts are reasonable.

The crucial point is that $\tau_\alpha \propto \alpha(T)/\Gamma$, so the amplitude of the ionizing background \emph{cannot} be measured independently of the temperature.  In practice, given the difficulty of independently measuring $T$ at the high redshifts of interest for reionization, studies typically take it to be constant both with time and density (e.g., \citealt{fan01,fan06}) or marginalize over it with a range of simple equations of state \citep{bolton07}.  Our models show that this is clearly too simple:  the \lya forest is most sensitive to $\Delta \sim 0.2$ at $z \sim 6$, where the temperature can vary by nearly an order of magnitude after reionization (see Fig.~\ref{fig:TH30}).  Naively, this leads to uncertainties in $\Gamma$ of a factor of a few.  Even more worrying are the systematic errors introduced by post-reionization cooling, which affect estimates of the \emph{evolution} of $\Gamma$.

\begin{figure}
\plotone{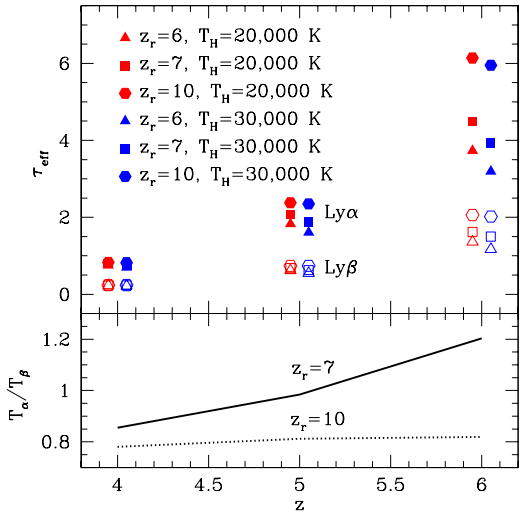}
\caption{\emph{Top panel:}  Effective optical depths for \lya (filled symbols) and Ly$\beta$ (open symbols) in three different reionization scenarios:  $z_r=6,\,7$, and 10 (triangles, squares, and hexagons, respectively).  In all cases, $\Gamma_{12}=0.1,\,0.2$, and 0.4 at $z=6,\,5,$ and 4, respectively.  The red and blue symbols take $T_H=20,000$ and $30,000 \kel$, respectively (note that they are slightly displaced from $z=6,\,5$, and 4 only for plotting purposes).  \emph{Bottom panel:}  Ratio of effective temperatures $T_{\alpha}/T_{\beta} = (A_{\alpha}/A_{\beta})^{0.7}$, for the models with $z_{r}=7$ and $10$ (solid and dotted curves, respectively), with $T_{\rm H}=30,000 \kel$, depicted in the top panel. }
\label{fig:taueff}
\end{figure}

The filled symbols in Figure~\ref{fig:taueff} give a practical example of this thermal evolution's effects on $\tau_{{\rm eff},\alpha}$.  Here we show how the transmission evolves for three reionization scenarios (with $z_r=6,\,7$, and 10 for the triangles, squares, and hexagons).  For concreteness, we take $\Gamma_{12}=0.1,\,0.2$, and 0.4 at $z=6,\,5,$ and 4 (chosen to match roughly the estimates from \citealt{fan06}).  The red and blue symbols take $T_H=20,000$ and $30,000 \kel$, respectively.

As expected, the differences between the three scenarios are small at $z=4$, because all the models give nearly the same temperature at that point (the difference is $\sim 5$ per cent in absolute transmission).  However, as we approach $z=6$, the transmission increases dramatically for the $z_r=6$ model:  it is over an order of magnitude larger than that of the $z_r=10$ model, purely because of the high temperatures achieved in low-density voids just after reionization.  This higher temperature helps to compensate for the increasing density of the Universe;  in other words, the rapid cooling following reionization slows the evolution of $\tau_{{\rm eff}, \alpha}$.  

The open symbols in Figure~\ref{fig:taueff} show the same quantity, but for the Ly$\beta$ transition.  This line has a significantly smaller oscillator strength (leading to a prefactor 6.24 times smaller in eq.~\ref{eq:fgpa}); it therefore samples higher densities, whose temperatures remain more stable after reionization.  As a result, $\tau_{{\rm eff},\beta}$ varies by a much smaller amount across these reionization models (differing by a factor of only $\sim 2.5$ in absolute transmission at $z=6$).

\begin{figure*}
\plottwo{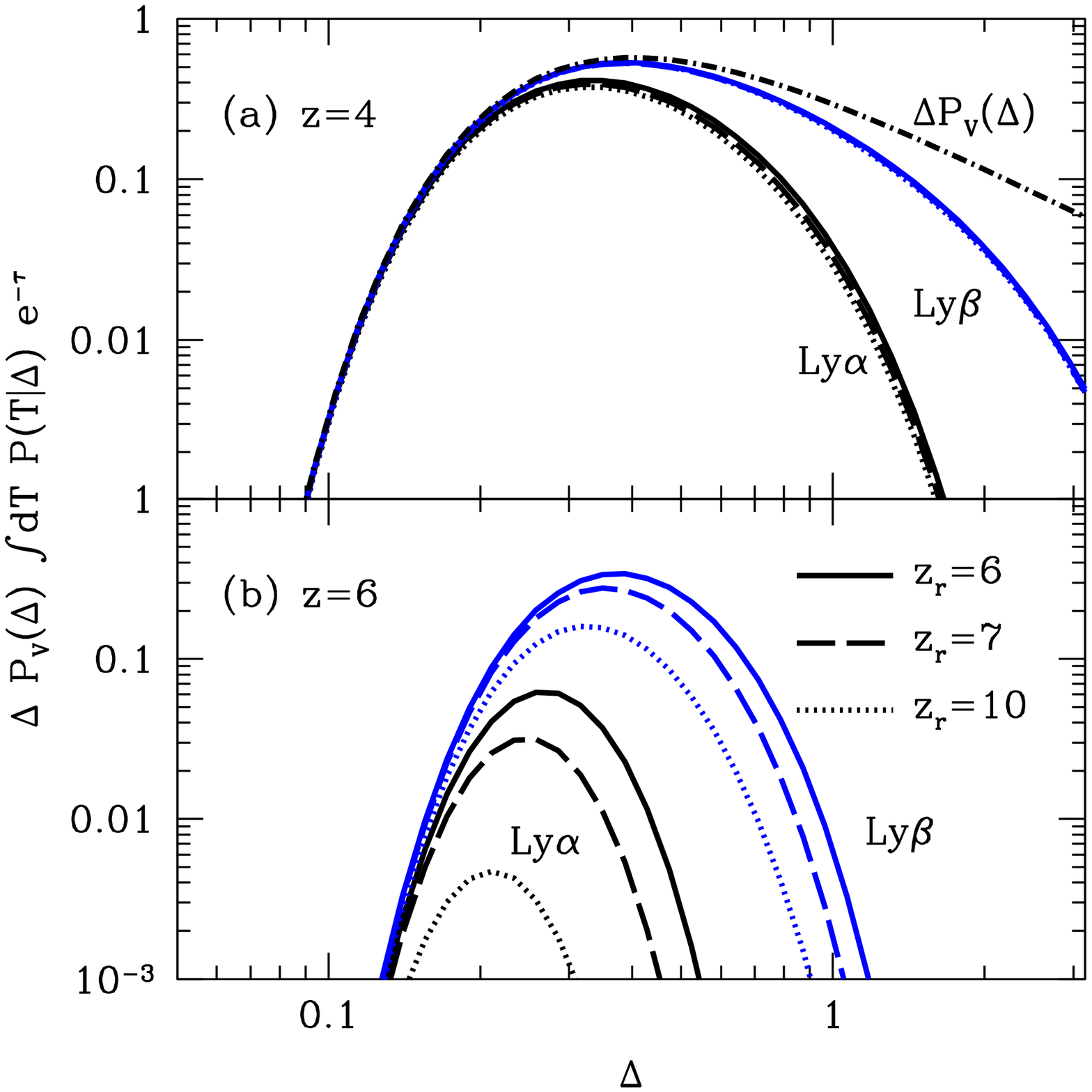}{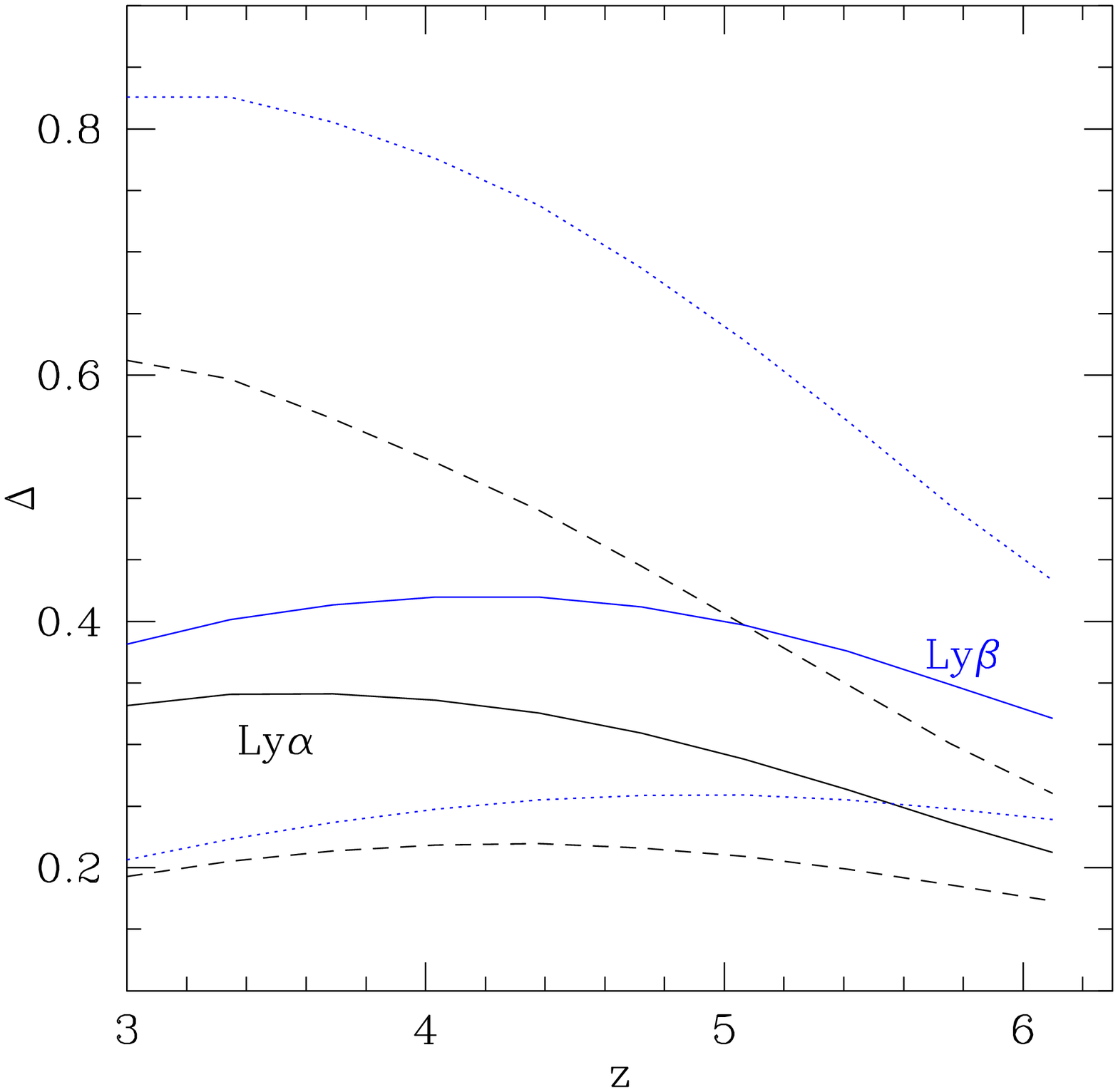}
\caption{\emph{Left panel:} Contribution by density to the total IGM transmission at $z=4$ (\emph{a}) and $z=6$ (\emph{b}).  The black curves are for \lya while the blue curves are for Ly$\beta$.  Within each set, the solid, dashed, and dotted curves assume reionization completes at $z_r=6,\,7,$ and $10$, respectively.  All curves assume $T_H=30,000 \kel$ and $\Gamma_{12}=0.1$ ($z=6$) and $0.4$ ($z=4$).  In the top panel, the dot-dashed curve shows the density distribution from eq.~(\ref{eq:pvd}).  \emph{Right panel:} Solid lines show the median overdensity which contributes to the total IGM transmission for the respective transitions (\lya and Ly$\beta$, bottom and top), while the dashed (dotted) lines show the overdensities between which 68\% of the flux transmission occurs for the respective transitions.}
\label{fig:lymanseries}
\end{figure*}

Figure~\ref{fig:lymanseries} provides some intuition about this behavior.  The left panel shows the logarithmic contributions by density to the outer integral in equation~(\ref{eq:taueff}), or in other words the range of densities that are actually visible in the forest.  The black curves are for \lya while the blue curves are for Ly$\beta$.  Within each set, the solid, dashed, and dotted curves assume that reionization completes at $z_r=6,\,7,$ and $10$, respectively.  All take $T_H=30,000 \kel$ and the same ionizing backgrounds as in Fig.~\ref{fig:taueff}. The right panel shows the median overdensity contributing to the total IGM transmission, as well as the overdensities between which 68\% of the flux transmission occurs. The figure assumes an isothermal equation of state and uses a spline fit to the data of \citet{songaila04} for ${\mathcal T}_\alpha$.

Several key points are obvious from the figure.  First, in the left panel all the curves converge at $\Delta \la 0.15$.  This is because voids with $\Delta \la 0.2$ are exponentially suppressed in the \citet{miralda00} density distribution and is a phenomenon independent of our temperature distributions.  On the other hand, the turnover at high densities is primarily due to the $e^{-\tau}$ factor and depends on both the transition and the temperature.  For example, the dot-dashed curve in (\emph{a}) shows $\Delta P_V(\Delta)$ at $z=4$.  All the other curves trace it closely until $\tau \sim 1$.  A second feature is also crucial: at $z=6$ the largest transmission occurs when $z_r=6$.  The order-of-magnitude increase in temperature opens up a substantial range of densities (over which $P_V$ is increasing rapidly), raising the transmission by a large factor. These points are amplified by the right panel, which shows the density dependence with redshift, given the the empirically observed Ly$\alpha$ flux transmission. While the median overdensity for Ly$\alpha$ and Ly$\beta$ transmission do not differ strongly, the latter transition accesses a wider range of densities above the median. On the other hand, the range of contributions for both transitions narrows significantly toward high redshift (when increasing densities and a weaker ionizing background imply that higher overdensities do not transmit flux -- leaving the observations heavily weighted toward voids). 

\section{Can We Measure the IGM Temperature at $z>5$?} \label{measure}

From a practical standpoint, the most important goal is to break the degeneracy between $\Gamma$ and the thermal structure.  In one sense, the narrow range of densities probed by the \lya forest is helpful, because it insulates us from uncertainties in the equation of state parameter $\gamma$:  the transmission in all of our models can be closely mimicked by choosing an isothermal equation of state with appropriate normalization $T$.  On the other hand, this same narrow range makes it even more difficult to measure evolution in the temperature, which shows up most clearly in the steepening of the equation of state with time. 

However, there is some hope, thanks to the higher-order Lyman transitions.  Figure~\ref{fig:lymanseries} shows that Ly$\beta$ is sensitive to a wider range of densities than \lyans, so it will effectively have a higher temperature at any redshift (and Ly$\gamma$, with an optical depth 18 times smaller than \lyans, will sample a still larger range).  Intuitively, measuring the effective temperature at several different densities through these different transitions provides a handle on the shape of the equation of state, which can then be combined with models like our own to break the $\Gamma$-$T$ degeneracy (and even to measure the reionization redshift from the evolving thermal properties, as in \citealt{theuns02-reion,hui03}).  

\citet{dijkstra04-lyb} first recognized that the Ly$\beta$ forest could be used to measure the equation of state, because it probes the IGM at higher overdensities relative to Ly$\alpha$ absorption. In particular, they showed that the small-scale Ly$\beta$ power spectrum was sensitive to the equation of state, whereas the Ly$\alpha$ power spectrum was relatively robust to such variations, due to its narrow density coverage. However, for our purposes it is not possible to measure the Ly$\alpha$ or Ly$\beta$ power spectra at high-redshift, given the paucity of data and the frequency of saturated absorption. Instead, we point out that the relative effective optical depths in Ly$\alpha$ and Ly$\beta$ absorption can already give us information about the equation of state. 

More formally, we have $\tau_{{\rm eff},i} = \tau_{{\rm eff},i}(A_i)$ for each transition $i=\alpha, \, \beta$, and $\gamma$, where $A_i = \Gamma/\alpha(T_i)$ and $T_i$ is the effective temperature for that transition.  As described above, these $T_i$ will differ because each transition probes a different range of densities.  On the other hand, $\Gamma$ is (naively) independent of the transition.  As in \citet{fan01, fan06}, we can then extract the $A_i$ and measure the \emph{slope} of the equation of state via $A_\alpha/A_\beta = (T_\alpha/T_\beta)^{0.7}$.  The redshift evolution of the slope can then reveal details of reionization and (after further modeling) $\Gamma$.  For example, the bottom panel of Figure~\ref{fig:taueff} shows the inferred ratio in two reionization scenarios, with $z_r=7$ and $z_r=10$.  Long after reionization, the temperatures no longer evolve, so $T_\alpha/T_\beta$ is nearly constant (as in the $z_r=10$ scenario).  But just after reionization, when the voids are cooling quickly, the ratio undergoes substantial evolution.  It is this effect that can constrain the timing of reionization.

For example, consider the line of sight to J1148+5251 (a $z=6.42$ quasar), which shows transmission in the Ly$\alpha$, Ly$\beta$, and Ly$\gamma$ troughs \citep{white03, oh05, white05}.  \citet{fan06} used a technique very similar to ours, but assuming an isothermal equation of state with $T=10^4 \kel$, to transform the amount of transmission into measurements of $\Gamma_{12}=0.06,\,0.04,$ and $0.015$ from the Ly$\alpha$, Ly$\beta$, and Ly$\gamma$ troughs, respectively.  Although the accompanying errors are large (and difficult to estimate) the discrepancies are certainly worrisome.  

We have argued that these should instead be considered measurements of $\Gamma T_i^{0.7}$, so that (assuming a constant $\Gamma$) the three estimates actually imply $T_\alpha \approx 2 T_\beta \approx 7 T_\gamma$ -- a \emph{strongly} inverted equation of state; indeed, this requires a much steeper density dependence ($\gamma - 1 \approx -2$) than seems reasonable (c.f. Fig.~\ref{fig:TH30}).  Preliminary calculations for other quasars in the $z\sim 4-6$ range also indicate that the majority of them have strongly inverted apparent equations of state. 
Clearly -- at least according to our models -- the post-reionization equation of state can account for only a small fraction of these differences.  Other effects may then be at play.

For example, $\Gamma$ may be much larger in low-density regions than in those probed by the higher-order transitions.  At first blush, this seems surprising, as the source density inside voids is much smaller than that in dense regions, so that $\Gamma_{\alpha}/\Gamma_{\beta} < 1$, making our derived $T_{\alpha}/T_{\beta}$ lower limits.  Of course, radiative transfer effects may help.  But the reduced abundance of Lyman limit absorbers in voids (which could increase the mean free path of ionizing photons and hence increase the radiation field) in unlikely to overcome the reduced abundance of galaxies (which are presumably more highly biased and hence more sensitive to the large scale density), in order to allow a net increase in $\Gamma$.  So one must appeal to more complex effects, such as shadowing.  For example, in the context of the helium-ionizing background, \citet{maselli05} found that radiative transfer hardens the background in low-density regions.  Whether a systematic increase in the ionizing background at lower overdensities can possibly account for the discrepancies remains unclear for now.

A comparison to the simulations of \citet{trac08} suggests another possibility:  increased scatter in the equation of state (particularly if it is density-dependent).  For a given mean equation of state $\bar{T}(\Delta)$, scatter increases flux transmission, much as an inhomogeneous density distribution does. However, the increase is {\it greater} for ${\mathcal T}_{\alpha}$ than for ${\mathcal T}_{\beta}$, implying that $\tau_{\rm eff, \alpha}/\tau_{\rm eff, \beta}$ will be biased downward, and hence $(T_{\alpha}/T_{\beta})$ will be biased upward. This is because scatter in the equation of state causes flux transmission to be less heavily weighted toward rare voids (since hot denser regions become just as transparent). The steepness of $P_V(\Delta)$ in the \lya transmission range (see Fig.~\ref{fig:lymanseries}) then implies a large increase in the total transmission.  Fluctuations in the ionizing background would have a similar effect. 

For instance, consider an isothermal mean equation of state ($\bar{T}$ and $\bar{\Gamma}$ are independent of density), where $A \equiv \Gamma_{-12} (T/20,000 \, {\rm K})^{0.7}$ has a lognormal distribution with $(\bar{A},\sigma_{\rm ln A})=(0.02,1)$, either due to density-independent scatter in the temperature or in the ionizing background. This gives $\tau_{\rm eff, \alpha}/\tau_{\rm eff, \beta}=1.9$ (as opposed to $\tau_{\rm eff, \alpha}/\tau_{\rm eff, \beta}=2.6$ for an isothermal equation of state without scatter), i.e. increased flux transmission in Ly$\alpha$ relative to Ly$\beta$. We would then infer $(T_{\alpha}/T_{\beta})=1.4$ from our naive approach -- an ``inverted" equation of state -- even though the gas is actually isothermal (in the mean). Such biases will be further amplified if the scatter increases toward lower densities, as in \citet{trac08}. Note, however, that a full evaluation of the impact of scatter requires numerical simulations, because some of the scatter in the low density gas may not be important. For example, the high temperatures in low-density gas near large scale overdensities may not manifest themselves in observations, since the absorption in such regions of the spectra is likely to be saturated anyway.   

Interestingly, both  scatter in the equation of state and any density dependence in the ionizing background should fall with time (the latter because the photon mean free path increases with time), and so all of these effects only increase the likelihood of an ``inverted" equation of state immediately after reionization, followed by a recovery toward the asymptotic relation. 

The above uncertainties notwithstanding, high-quality spectra of the sort already available might still provide some useful constraints.  As an example, suppose that we have measured $\tau_{{\rm eff},\alpha}=6.1$ at $z=6$.  Models with $z_r=6$ and $\Gamma_{12}=0.0385$ or $z_r=10$ and $\Gamma_{12}=0.1$ both provide this level of \lya absorption, but $\tau_{{\rm eff},\beta}=2.34$ and $2.07$ for the respective scenarios (or a $\sim 20\%$ difference in transmitted flux).  How many lines of sight are required to distinguish between these predictions?  There are four sources of uncertainty on $\tau_{{\rm eff},\beta}$:  cosmic variance (which provides a \emph{maximal} fractional 
error $\sim 0.4$ on the Ly$\beta$ transmission for this value of opacity; \citealt{lidz06}),\footnote{Actually, this value corresponds to the cosmic variance on different lines of sight.  We are comparing the \lya and Ly$\beta$ forests along the \emph{same} line of sight, so the cosmic variance should be much smaller -- sourced only by the decorrelation between Ly$\alpha$ and Ly$\beta$ from sampling different densities.  This aspect must be calibrated with simulations.} uncertainty in the foreground \lya absorption ($\tau_{\rm fg}=2.38 \pm 0.32$; \citealt{oh05}), uncertainties in estimating the quasar continuum $F_Q$ ($\sim 10\%$ at Ly$\beta$; \citealt{oh05}), and measurement error (which is much smaller than the other uncertainties for high-quality spectra; e.g., \citealt{white03}).  Thus the total uncertainty in the measured flux $F_{\rm obs}$ over $N$ lines of sight is 
\bqa
\delta F_{\rm obs} & \approx & {F_{\rm obs} \over \sqrt{N}} \sqrt{(\delta \tau_{\rm fg})^2 + (\delta \tau_{\rm cv})^2 + \left({\delta F_Q \over F_Q} \right)^2} \label{eq:error} \\ 
& \approx & 0.55/\sqrt{N},
\nonumber
\eqa
where the second term in the square root conservatively accounts for cosmic variance.  Thus, only $\sim 5$ (20) lines of sight would be needed to distinguish these two reionization scenarios at 68\% (95\%) confidence -- provided that the \lya transmission can itself be cleanly estimated (and provided we can robustly model the spatial distribution of $\Gamma$ and the temperature scatter).  Working with \lya is somewhat harder as absorption is more easily saturated, yielding only lower bounds on $\tau_{\rm eff, \alpha}$, and hence upper bounds on $T_{\alpha}/T_{\beta}$. However, our results show that,
with improved modeling,
constraints on the timing of reionization should be possible. 

\section{Discussion} \label{disc}

We have described how inhomogeneous hydrogen reionization affects the thermal structure of the IGM.  Because it proceeds ``inside out," from high to low density regions \citep{barkana04, furl04-bub}, voids are ionized last and so are \emph{hotter} than denser gas near the end of reionization (see also \citealt{bolton04, trac08}).  Rapid adiabatic and Compton cooling work quickly to erase this inversion, and the IGM equation of state passes through a nearly isothermal phase (at $\Delta \la 1$) and then approaches its relatively steep and monotonic asymptote.  Late reionization at $z_r=6$ leaves the low-density gas in the isothermal phase at $z \sim 4$; if reionization is earlier, it is close to the asymptotic form.  In our model, the scatter around the median temperature at low and high densities is relatively small, because extreme densities are all either ionized very near the end of reionization (for $\Delta \ll 1$) or very near the beginning.  The scatter is larger near the mean density, because the spread in reionization redshifts is much larger.  The recent simulations of \citet{trac08} show larger scatter at high and low densities, perhaps indicating that radiative transfer plays a significant role in the temperature distribution.

Although similar trends may occur during helium reionization at $z \sim 3$ (FO08),``inside out" reionization is less relevant  in practice during that phase, because the ionizing sources are so rare and bright that the correlation between large-scale overdensities and ionized regions is much weaker and because radiative transfer is much  more complex during this era \citep{furl08-helium, mcquinn08, bolton08}.  From a theoretical standpoint, an ``inside out" picture is much more likely to be accurate during hydrogen reionization, which is driven by large groupings of small sources whose random fluctuations are small \citep{furl05-charsize}.  

The \lya forest is so thick at $z \ga 5$ that individual lines can no longer be distinguished, so the techniques used to measure the IGM temperature directly at lower redshifts \citep{schaye00, ricotti00, mcdonald01} are no longer useful.  However, the thermal structure does still affect the forest:  we found that shifting the end of reionization over the range $z_r=6$--10 can change the total transmission by up to an order of magnitude at $z=6$, although by $z=4$ these differences have nearly disappeared (thanks largely to the decreasing mean density of the IGM, which makes a much larger fraction of the IGM visible).  

This raises two important questions for understanding the high-$z$ IGM.  First, can the temperature evolution be measured and used to constrain the timing of reionization?  We have shown that the expected variation in temperature is modest at $z \la 4$, so using that era is difficult (even leaving aside possible contamination from helium reionization; \citealt{theuns02-reion, hui03}).  Moreover, observations yield significantly higher temperatures than any published model, with errors much larger than the differences between the models \citep{schaye00, zald01}.  At higher redshifts, degeneracy with the amplitude of the ionizing background is troublesome:  an isothermal IGM is completely degenerate with $\Gamma$, and, for the \lya transition at $z \ga 4$, isothermality is typically an excellent approximation, because the forest is only able to sample a very limited range in densities.  

However, higher-order transitions can help significantly (at least in principle), because they sample a wider range of densities (see Fig.~\ref{fig:lymanseries}) and hence a different portion of the equation of state.  Deviations from isothermality can be measured via the transmission ratios and used to map the equation of state (and hence reionization redshift).  This is complicated by the rapidly diminishing transmission as well as contamination from the lower-redshift \lya forest, but we have found that differentiating $z_r=6$ and $z_r=10$ via the Ly$\beta$ transmission only requires a modest increase in the number of known lines of sight to distant quasars.  However, we will certainly need more sophisticated models that include variations in the ionizing background across these different environments, as well as scatter in the equation of state.

The second consequence is for measurements of $\Gamma$, which are particularly interesting near reionization.  To date, claimed constraints on $\Gamma$ at $z=6$ have ignored thermal evolution, either assuming isothermality (e.g., \citealt{fan01, fan06}) or marginalizing over standard equations of state \citep{bolton07}.  Here we have shown that the evolution is much more complex, which will systematically affect measurements of the ionizing background.  Unfortunately, with existing data it is not at all clear how these effects will play out.

As shown in Figure~\ref{fig:taueff}, just after reionization both the \lya and Ly$\beta$ optical depths evolve quite slowly, even if $\Gamma$ is increasing, because rapid IGM cooling cancels out much of the change in $\Gamma$ as well as the overall density evolution.  If reionization occurs late, $\Gamma$ must therefore evolve even \emph{more} quickly than commonly assumed over $z \sim 5$--$6$:  in our models, if reionization ends at $z=6$, so that $\VEV{T_\alpha} \sim 25,000 \kel$, we would require $\Gamma_{12} \approx 0.04$ to attain $\tau_{{\rm eff},\alpha} \approx 6$ (as opposed to $\Gamma_{12}=0.1$ if $z_r=10$).  Thus, if the conventional explanation is correct, the ionizing background would have to increase by a factor of five from $z=6$ to $z=5$; whether this is reasonable depends entirely on the presumed evolution of Lyman-limit systems, which is highly uncertain \citep{furl08-mfp}.  If instead reionization -- or at least reheating -- ends much earlier, $\Gamma$ could increase relatively slowly over the same time interval.

Without more detailed modeling, for now it is perhaps better to consider uncertainties in the equation of state to be ``noise" in measurements of the ionizing background.  Our model suggests that such systematic uncertainties can be relatively large, even compared to the substantial statistical errors on the existing observations; for example,  \citet{bolton08} estimate $\Gamma_{12}(T_{\alpha}/10,000 \kel)^{0.7}=0.2^{+0.15}_{-0.2}$ at $z=6$.  Allowing $\VEV{T_\alpha}$ to vary from $4,000$--$25,000 \kel$ increases the allowed range to $\Gamma_{12} = 0$--$0.6$.  Thus, even today's constraints need to account for the thermal effects of reionization.

It is important to note that our model neglects (at least) three aspects of photoheating that will affect the \lya forest, so more detailed simulations will be required for careful comparison to observations.  First, our fluctuating Gunn-Peterson approximation (used in \S \ref{lyaforest}) ignores thermal broadening and does not perfectly reproduce the line structure of the forest \citep{bolton05, tytler04, jena05}.  For example, the heating that accompanies reionization will make the forest lines broader, decreasing the number of saturated lines and increasing the overall optical depth.  This will oppose the direct effect of heating (which reduces the neutral fraction) and so partially compensate for the decreased optical depth following reionization.  As the Universe cools, the lines will become narrower and more saturated, again opposing the slow decrease in the overall transmission in our results.  This will make photoheating even more difficult to observe. 

Second, we have ignored radiative transfer effects during reionization, where the radiation field hardens as it propagates outward from an ionizing source \citep{abel99}. Note that such spectral hardening effects could also lead to a systematic tilt in the initial equation of state, in that voids may see a more heavily filtered and hardened radiation field, and hence be heated to higher temperatures \citep{bolton04}. Such effects may stymie efforts to infer the topology of reionization from an inverted equation of state. However, we expect them to be much weaker for hydrogen reionization as compared to helium reionization, since the former case is driven by a softer (primarily stellar) radiation field. Furthermore, the shape of the UV radiation field does not significantly affect the post-reionization evolution of the equation of state: the relatively low recombination rate of hydrogen means that the gas largely loses thermal contact with the radiation field. Thus, efforts to infer the redshift of reionization using the techniques we have discussed may still be robust to radiative transfer effects, although this requires more detailed study with numerical simulations.  Other than the increased scatter, comparison to the simulations of \citet{trac08} does not reveal any substantial biases in our approach.

The final effect is more subtle:  the increased temperature following reionization will also increase the pressure of the IGM, leading to a rearrangement of IGM gas through Jeans smoothing.  While well-understood in the context of minihalos (where it is called ``photoevaporation"; \citealt{shapiro04, iliev05-mh}), the analogous IGM processes have been considered only recently \citep{pawlik08}.  Jeans smoothing will move material away from high-density regions, ``puffing out" filaments to fill voids and making $P_V(\Delta)$ more strongly peaked around the mean density.  \citet{pawlik08} found that this adjustment occurs over approximately an expansion time (roughly the sound-crossing time at densities near the cosmic mean).  This will affect both the mean transmission [through $P_V(\Delta)$] and the mean free path of ionizing photons (which depends on dense clumps; \citealt{furl05-rec}).  However, our simple analytic model does not suffice to describe this process, so we defer its inclusion to future, more detailed work. 

Nonetheless, it is worth noting that photoheating reduces transmission at late times once these hydrodynamic effects set in, making the voids, which are the primary source of transmission, higher density and so more opaque. As redshift decreases, this creates a systematic bias toward lower and lower estimates of $\Gamma$ compared to the true value, just like the temperature evolution effects we have been discussing.  If so, $\Gamma$ must evolve even faster than we have quoted above. Such effects also reduce ${\mathcal T}_{\alpha}/{\mathcal T}_{\beta}$ (since ${\mathcal T}_{\alpha}$ is more heavily weighted toward transmission in the voids) and by extension the inferred $T_{\alpha}/T_{\beta}$, which would exacerbate our difficulty explaining the measured ratio. On the other hand, in the context of helium reionization, similar hydrodynamic effects have a relatively small imprint on $\tau_{\rm eff}$ \citep{bolton09}. 

\acknowledgments

This research was partially supported by the David and Lucile Packard Foundation (SRF), grant NSF-AST-0607470 (SRF), and NASA grant NNG06GH95G (SPO). 


\end{document}